\documentclass[aps,preprintnumbers,superscriptaddress,longbibliography,amsmath,amssymb,twocolumn,10pt,floatfix]{revtex4-1}

\usepackage{xr}
\usepackage{amssymb}
\usepackage{multirow}
\usepackage{siunitx}
\usepackage{makecell}

\usepackage{comment} 
\usepackage{graphicx}
\usepackage{xspace}
\usepackage{siunitx}
\usepackage{xfrac}
\usepackage[utf8]{inputenc}
\usepackage{dcolumn}
\usepackage{xcolor}
\definecolor{darkred}{rgb}{0.3,0,0}
\definecolor{darkblue}{rgb}{0,0,0.3}
\definecolor{firebrick}{rgb}{0.5,0.125,0.125}
\definecolor{darkgreen}{rgb}{0,0.3,0}
\usepackage[colorlinks=true,linkcolor=firebrick,citecolor=darkgreen,urlcolor=darkblue]{hyperref}

\usepackage{lineno}

\usepackage[caption=false]{subfig}

\newcommand*{\addFileDependency}[1]{
\typeout{(#1)}
\IfFileExists{#1}{}{\typeout{No file #1.}}
}\makeatother

\newcommand*{\myexternaldocument}[1]{%
\externaldocument{#1}%
\addFileDependency{#1.tex}%
\addFileDependency{#1.aux}%
}

\myexternaldocument{supplement}

\begin{document}




\title{The Potential of Water-Cherenkov Air Shower Arrays for detecting transient sources of high-energy astrophysical neutrinos}

\date{\today}

\author{J.~Alvarez-Mu\~niz}
\affiliation{Instituto Galego de F\'\i{}sica de Altas Enerx\'\i{}as (IGFAE), Universidade de Santiago de Compostela, Santiago de Compostela, Spain}

\author{R.~Concei\c{c}\~ao}
\affiliation{Laborat\'orio de Instrumenta\c{c}\~ao e F\'\i{}sica Experimental de Part\'\i{}culas -- LIP and Instituto Superior T\'ecnico -- IST, Universidade de Lisboa -- UL, Lisboa, Portugal}

\author{P.~J.~Costa}
\affiliation{Laborat\'orio de Instrumenta\c{c}\~ao e F\'\i{}sica Experimental de Part\'\i{}culas -- LIP and Instituto Superior T\'ecnico -- IST, Universidade de Lisboa -- UL, Lisboa, Portugal}

\author{B.~S.~Gonz\'alez}
\affiliation{Laborat\'orio de Instrumenta\c{c}\~ao e F\'\i{}sica Experimental de Part\'\i{}culas -- LIP and Instituto Superior T\'ecnico -- IST, Universidade de Lisboa -- UL, Lisboa, Portugal}

\author{M.~Pimenta}
\affiliation{Laborat\'orio de Instrumenta\c{c}\~ao e F\'\i{}sica Experimental de Part\'\i{}culas -- LIP and Instituto Superior T\'ecnico -- IST, Universidade de Lisboa -- UL, Lisboa, Portugal}

\author{B.~Tom\'e}
\affiliation{Laborat\'orio de Instrumenta\c{c}\~ao e F\'\i{}sica Experimental de Part\'\i{}culas -- LIP and Instituto Superior T\'ecnico -- IST, Universidade de Lisboa -- UL, Lisboa, Portugal}

\begin{abstract}
We highlight the capacity of current and forthcoming air shower arrays using water-Cherenkov stations to detect neutrino events spanning energies from $10\,$GeV to $100\,$TeV. This detection approach leverages individual stations equipped with both bottom and top photosensors, making use of features of the signal time trace and machine learning techniques. Our findings demonstrate the complementary of this method to established and future neutrino-detection experiments, including IceCube and the upcoming Hyper-Kamiokande experiment.
\end{abstract}

\maketitle

\section{Introduction}
\label{sec:intro}

Transient emissions originating from astrophysical sources can release a substantial burst of energy over brief time periods. These emissions may take the form of isolated events that have not been previously observed, or manifest as abrupt surges in the activity of well-known astrophysical sources. Several new categories of transients have been discovered through the observation of electromagnetic and/or gravitational radiation. These include a wide range of astrophysical phenomena, such as X-ray pulsars, Magnetars, Supernovae, Hypernovae, Gamma-Ray bursts (GRB), mergers of compact binary objects like Binary Neutron Stars (BNS) and Binary Black Holes (BBH), Tidal Disruption Events (TDE), AGN blazars, and more~\cite{Guepin:2022qpl}. This diversity highlights the large amount of astrophysical sources that can provide insights into the most violent and energetic processes in the universe. The burst in electromagnetic or gravitational emission can be associated with the acceleration of charged cosmic rays. Particle interactions within the source can result in the generation of secondary neutrinos and gamma rays, particularly in the range from GeV to PeV \cite{Ahlers:2015lln,Ackermann:2022rqc,Kurahashi:2022utm}. 

Large ground-based water-Cherenkov detector (WCD) arrays have been and are being used for high-energy gamma ray detection in the few tens of GeV to PeV energy range \cite{HAWC,LHAASO,SWGO}. These detectors offer advantages over Imaging Cherenkov Telescope Arrays as they operate almost continuously with close to a 100\% duty cycle and can monitor a substantial portion of the sky simultaneously. In this work, we demonstrate that such arrays can also effectively detect high-energy neutrinos, complementing dedicated neutrino experiments and contributing to the multimessenger observation of transient events.

Identification of upward-going particles in a WCD unit would be a clear signature of shower and/or muon events produced by GeV-PeV neutrinos crossing the Earth, with no known background. To explore this, we have combined cuts on observables accessible in WCD stations with modern machine-learning techniques, exploiting the time information recorded by the photo-multiplier tubes (PMT) in the WCD station. 
In this work, we show that for certain photo-sensors configurations in the station, it is possible to determine that particles are travelling in the upward direction. This allows the rejection of  the overwhelming downward-going particle background from conventional cosmic-ray atmospheric showers. Additionally, we demonstrate that the directional reconstruction accuracy would allow the identification of a relatively small solid angle in the sky, facilitating the detection of potential transient neutrino sources. This capability enables the investigation of regions in the sky where transient events are obscured from gamma-ray detection due to the Earth, but are detectable in high-energy neutrinos. The resolution for neutrino energy reconstruction is however expected to be rather low. In this regard, the WCD detector array is primarily expected to act as a neutrino counter above a certain energy threshold.

The total water volume of these WCD arrays, comprising thousands of individual detectors, would enable the observation of a significant detection rate, thereby complementing existing neutrino observatories.
The atmospheric neutrino flux in the TeV range would \textit{a priori} dominate the event rate. However, the detection of a sufficiently intense flux of high-energy neutrinos from a transient source is experimentally favoured because time-dependent excesses of events can stand above the atmospheric neutrino background. Follow-up of electromagnetic or gravitational wave alerts can further increase the statistical significance of potential associations of neutrinos with their sources.

An array of WCD could operate in this neutrino detection mode in real time with almost 100\% duty cycle, simultaneously with the gamma-ray detector mode, serving as both a follow-up and alert observatory of transient events. This novel neutrino operation mode has the potential to extend the capabilities and scientific scope of ground-based air-shower arrays primarily dedicated to gamma-ray detection. This potential is particularly important in the emerging field of Multimessenger Astronomy with high-energy particles.

\section{Experimental configuration and analysis strategy}
\label{sec:analysis}

In this work, we will be considering the layout of a gamma-ray wide field-of-view gamma-ray observatory, such as the planned SWGO~\cite{SWGO}. One of the proposed configurations involves a compact array of WCD stations, covering an area of $80\,000\,{\rm m^2}$ with a fill factor (FF)\footnote{Fill factor represents the portion of the detector area equipped with instrumentation in relation to the total coverage area.} of approximately $80\%$. This compact array is to be complemented with a sparser array, spanning an area of $\gtrsim 1\,{\rm km^2}$ with a FF of around $5\%$. The compact array is specifically designed for detecting low-energy showers (around $100\,$GeV), while the sparse array is tailored to capture ultra-high-energy gamma rays (up to a few PeV).

\begin{figure}[!h]
 \centering
\includegraphics[width=0.3\textwidth]{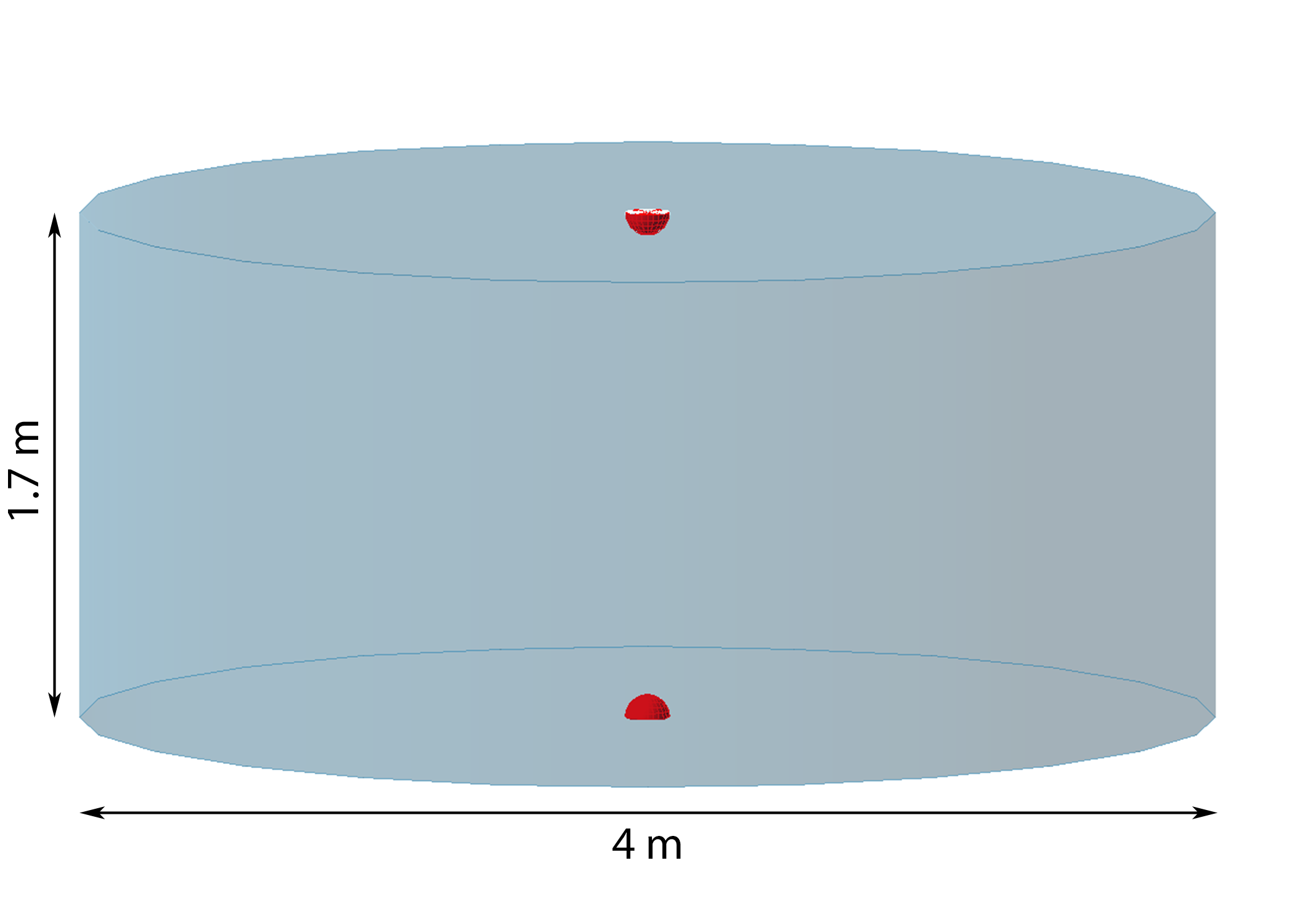}%
 \caption{Design concept of the water-Cherenkov detector (WCD) used in this study, with the red domes indicating the position and size of the photosensors.}
 \label{fig:WCDs:M1T1}
\end{figure}

The WCD unit used in this work is based on the designs presented in~\cite{WCD4PMTs} and~\cite{wcd2022mercedes}. While the dimensions of the cylindrical WCD station remain the same in both designs ($1.7\,$m height and $\approx 12\,{\rm m^2}$ in the base), the number and position of the photosensors are changed, reduced to only two. 
The interior walls of the WCD are covered with a reflective material to enhance the collection of Cherenkov light.

Given the dimensions of the WCD and the array layout configuration described above, such an experiment would have $5720$ stations in the compact array and $3660$ stations in the sparse array.

The WCD is equipped with two 8-inch PMTs positioned at the center of the cylinder caps, one at the bottom and another at the top (see Fig.~\ref{fig:WCDs:M1T1}).
Such configuration allows the detection of the direct Cherenkov light that hits the PMTs without being reflected on the inside walls of the WCD, maximizing the possibility of detecting the direct Cherenkov light pulse, regardless of whether the particles in the shower enter the water volume from above or below. The comparison between the time trace of the signal of the PMT hit by direct light with the opposite PMT, which should observe no direct Cherenkov light, can be used to tag the direction of the particles crossing the station. Additionally, the PMT in the top offers redundancy and could be used with a different dynamical range with respect to the bottom one to tackle potential saturation effects in showers with large signals at the ground.

\begin{figure}[!h]
 \centering
  \subfloat{
   \label{fig:mean_trace:down-going}
    \includegraphics[width=0.23\textwidth]{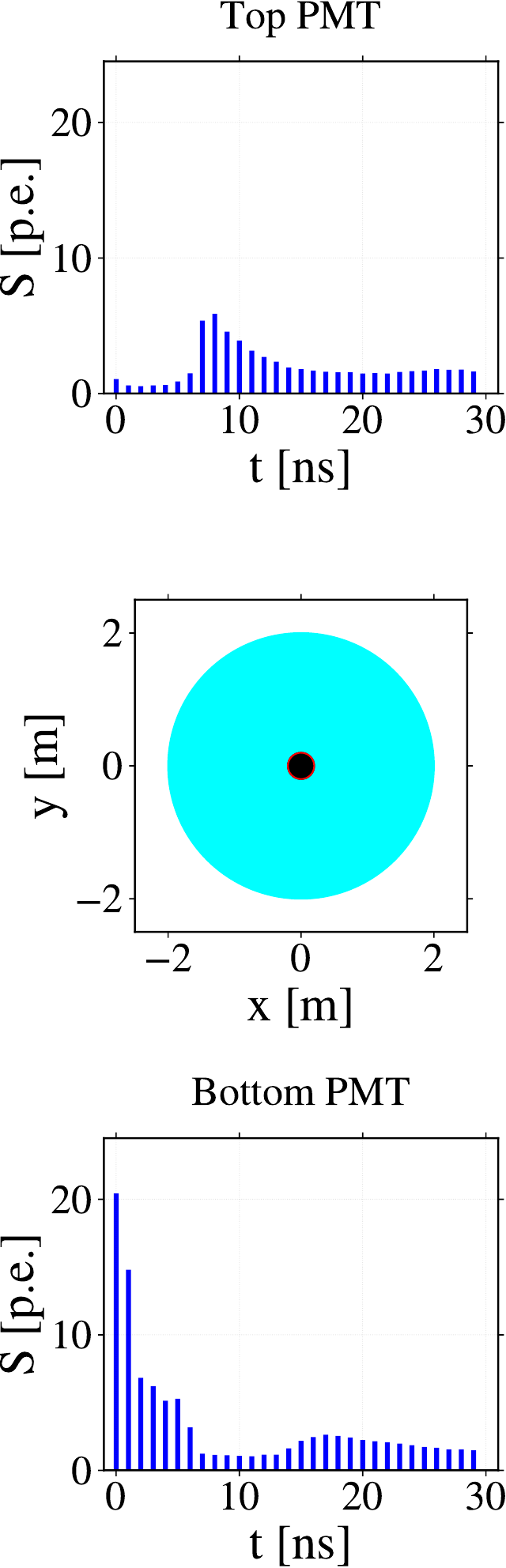}}
\hspace{0.001in}
  \subfloat{
   \label{fig:mean_trace:up-going}
    \includegraphics[width=0.23\textwidth]{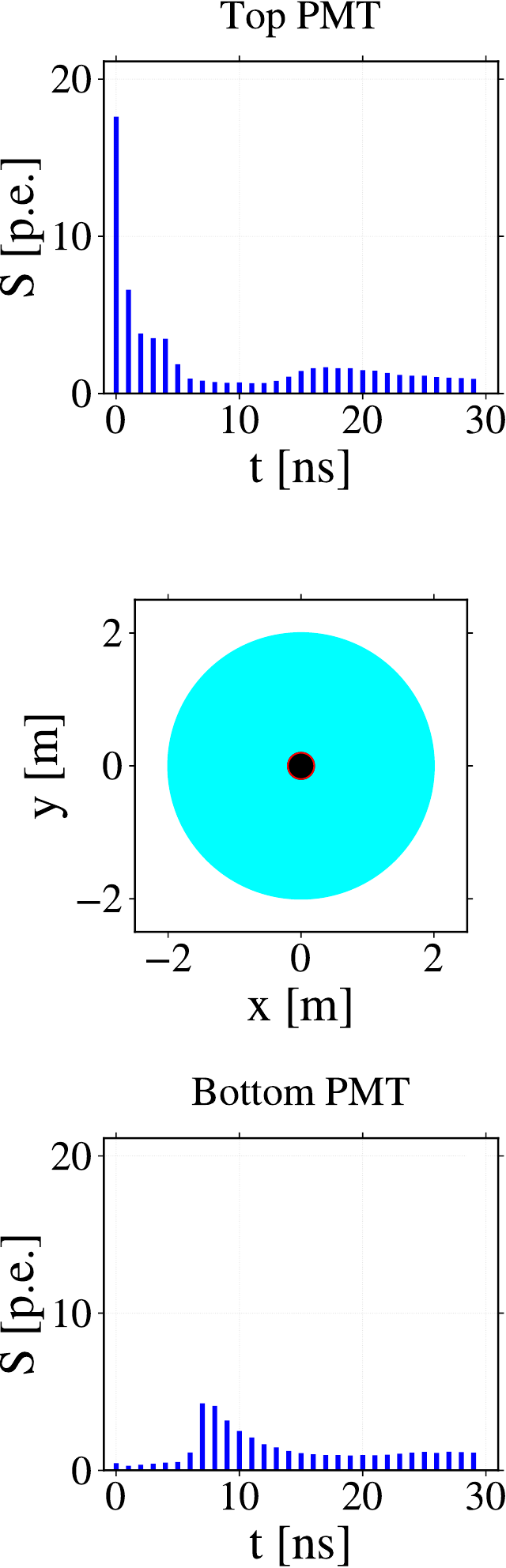}}
 \caption{Mean signal time traces in the top and bottom PMTs of the WCD considered in Fig.\,\ref{fig:WCDs:M1T1} for downward-going events with $\theta \leq 5^{\circ}$ (left panel), and upward-going events with $\theta \geq 175^{\circ}$ (right panel). The mean traces were computed using individual electrons, muons and protons with energies of $1\,$GeV, $2\,$GeV and $10\,$GeV respectively.}
 \label{fig:mean_traces}
\end{figure}

The features of the direct light as observed in the WCD mentioned above are evident in the simulations performed with Geant4~\cite{agostinelli2003geant4,Geant4_2016} shown in Fig.~\ref{fig:mean_traces}, where the average PMT signal time traces are displayed for both downward and upward events. Direct light, arriving earlier than the reflected one, and appearing in the first time bins of the traces, can be seen in the bottom (top) PMT for downward-going (upward-going) particles.

While the idea behind this analysis is simple, the complex geometries of particles entering a WCD unit, when crossed by an upward-going neutrino-induced shower, require more advanced analysis techniques to identify them.

\begin{figure*}[!htb]
 \centering
\includegraphics[width=0.9\linewidth]{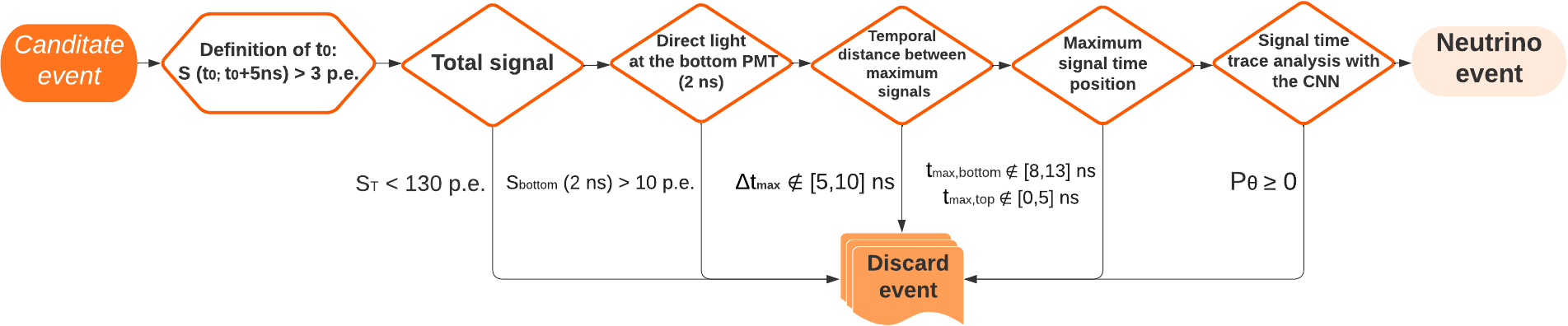}
\includegraphics[width=0.9\linewidth]{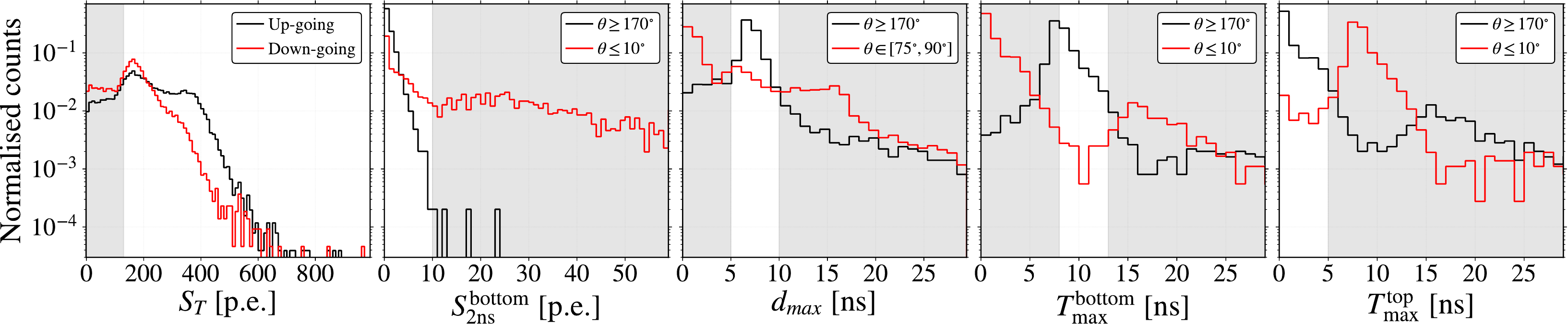}
\caption{(top) Steps followed in the analysis to identify upward-going neutrinos. (bottom) Distributions of several observables derived from the signal traces registered by the top and bottom PMTs, and cuts to suppress the downward-going background for the WCD in Fig~\ref{fig:WCDs:M1T1}. The shaded gray bands indicate the regions excluded by the cut in each corresponding variable. From left to right: Threshold in the total number of photoelectrons, $S_T$, obtained from muonic events to avoid clipping trajectories and low-energy background particles; signal collected during the first $2\,$ns at the bottom PMT, $S_{2\,\rm ns}^{\rm bottom}$, for vertical upward (black) and down-going events (red); difference between the times of the peaks of the signals in the top and bottom PMTs, $d_{\rm max}$, for close to horizontal (red) and upward-going events (black); time position of the peak of the signal at the bottom PMT, $T_{\rm max}^{\rm bottom}$, for vertical upward (black) and down-going events (red); time position of the peak of the signal at the top PMT, $T_{\rm max}^{\rm top}$, for vertical upward (black) and down-going events (red). See Fig.\,\ref{fig:Ptheta} for the performance of the Convolutional Neural Network (CNN).}
\label{fig:scheme_model} 
\end{figure*}

A dedicated simulation~\cite{wcd2022mercedes} was conducted to evaluate the discrimination capabilities of the WCD unit to particle showers induced by upward-going neutrinos with zenith angle $\theta \in [140^\circ,180^\circ]$, and background particles with $\theta \in [0^\circ,90^\circ]$. The simulation is focused on the aforementioned compact array and was implemented using the Geant4 toolkit (version 4.10.05.p01)~\cite{agostinelli2003geant4,Geant4_2016}. The conversion of Cherenkov photon hits into photoelectrons, which are subsequently binned in $1\,$ns time intervals, follows the methodology employed in~\cite{LATTES,WCD4PMTs,wcd2022mercedes}.
 
To optimize the selection of upward-going particles through the detection of direct Cherenkov light in the top PMT, we targeted only simulated upward-going showers above a zenith angle threshold. Given the angle of Cherenkov light in water, which is approximately $41^\circ$, we have conservatively selected in the simulation only particles with $\theta > 140^\circ$, to ensure that the bottom PMT is not directly hit by Cherenkov light.

For the purposes of this study, the primary sources of background were assumed to be showers induced by cosmic protons and atmospheric muons~\cite{pdg}\footnote{Note that at the evaluated altitudes, the number of secondary protons reaching the ground is comparable to the number of atmospheric muons.}. The choice of proton-induced showers provides a conservative estimate, as heavier elements are scarcer at lower energies. Furthermore, showers induced by heavier elements develop higher in the atmosphere, fluctuate less, and consequently are easier to discriminate.
The energy of the simulated vertical proton-induced showers ranges from $100\,$GeV to $1\,$TeV. These showers were simulated with the CORSIKA air shower Monte Carlo (version 7.5600)~\cite{CORSIKA}.
To assess the response of the detector to atmospheric muons as well as electrons and protons produced in small showers, these were injected in the WCD to study its response for various geometries.

The interaction of electron- and muon-neutrinos in the Earth's surface was simulated using HERWIG (version herwig6521)~\cite{HERWIG}. The directions of the resulting secondary particles were then appropriately rotated to move in the upward direction, directed towards the WCD. The neutrino-induced shower was subsequently simulated using Geant4. This comprehensive simulation considers ground density ($\rho=2.8\,{\rm{g\,cm^{-3}}}$) and its specific material properties, accounting for all relevant physics processes in the shower.

A basic local WCD trigger was implemented to replicate a realistic experimental analysis. The starting time of the traces collected at the WCD station, $T_0$, is defined as the time at which at least $5$ photo-electrons have been recorded by both PMTs.

Using the simulations described above, an analysis was conducted at the single WCD station level to determine optimal geometries and selection criteria for the identification of upward-going neutrino events. This analysis was complemented with a machine learning algorithm exploiting the features of the time trace of the PMT signals. The essential stages of this analysis are shown schematically in Fig.~\ref{fig:scheme_model}. The main objective of the selection criteria is to ensure that no background events originating from atmospheric muons or downward-going air showers is retained, thereby exclusively selecting upward-going neutrino events.
The series of cuts, shown and described in detail in the bottom panel and caption of Fig.\,\ref{fig:scheme_model}, effectively eliminate $\approx 99\%$ of the background events. These cuts are applied prior to the implementation of the machine learning algorithm, a Convolutional Neural Network (CNN) in this work.

Firstly, the cuts applied on the total signal ($1^{\rm st}$ and $2^{\rm nd}$ panel at the bottom of Fig.\,\ref{fig:scheme_model}) serve to eliminate \textit{clipping} particles that have a small track inside the WCD as well as low-energy background particles, including muons undergoing decay inside the WCD, thermal neutrons, and natural radioactivity. 
The cuts applied on the features of the PMT signal time trace, allow us to reject specific geometries that have been observed to hinder the discrimination. For example, configurations involving horizontal particles are excluded because Cherenkov light might hit both PMTs nearly simultaneously.
The chosen simulated upward-going particle trajectories exhibit a comparably narrow distribution of the time separation between the maxima of the top and bottom PMT signal traces ($d_{\rm max}$ in the $3^{\rm rd}$ panel at the bottom of Fig.\,\ref{fig:scheme_model}) that contributes to the discrimination of upward-going particles. Also, the times themselves (relative to $T_0$) of the maximum signals in the top and bottom PMT ($T_{\rm max}^{\rm top}$ in the $4^{\rm th}$, and $T_{\rm max}^{\rm bottom}$ in the $5^{\rm th}$ panel of Fig.\,\ref{fig:scheme_model}) were found to contain information on the direction of propagation of the particles.

Despite the prescribed cuts, a small fraction of simulated background events remained. Closer examination of the PMT signal time traces revealed additional features to distinguish between downward and upward-going trajectories. To address this in a highly-efficient manner, a CNN was employed for the analysis of the PMT traces following a similar approach as in \cite{4PMTs_NCA,WCD4PMTs,wcd2022mercedes}. The CNN was trained using a dataset that included $1\,$GeV electrons, $2\,$GeV muons, and $10\,$GeV protons and the interaction products of $10\,$GeV neutrinos. These particles were injected into the WCD station with random azimuthal angles, $\phi \in \left[0^{\circ},360^{\circ}\right]$, and zenith angles, $\theta \in \left[0^{\circ},180^{\circ}\right]$. The CNN weights were carefully adjusted to give a stronger penalty for misclassifying background events rather than signal (upward-going) events.

\begin{figure}[htb]
 \centering
s\includegraphics[width=0.9\linewidth]{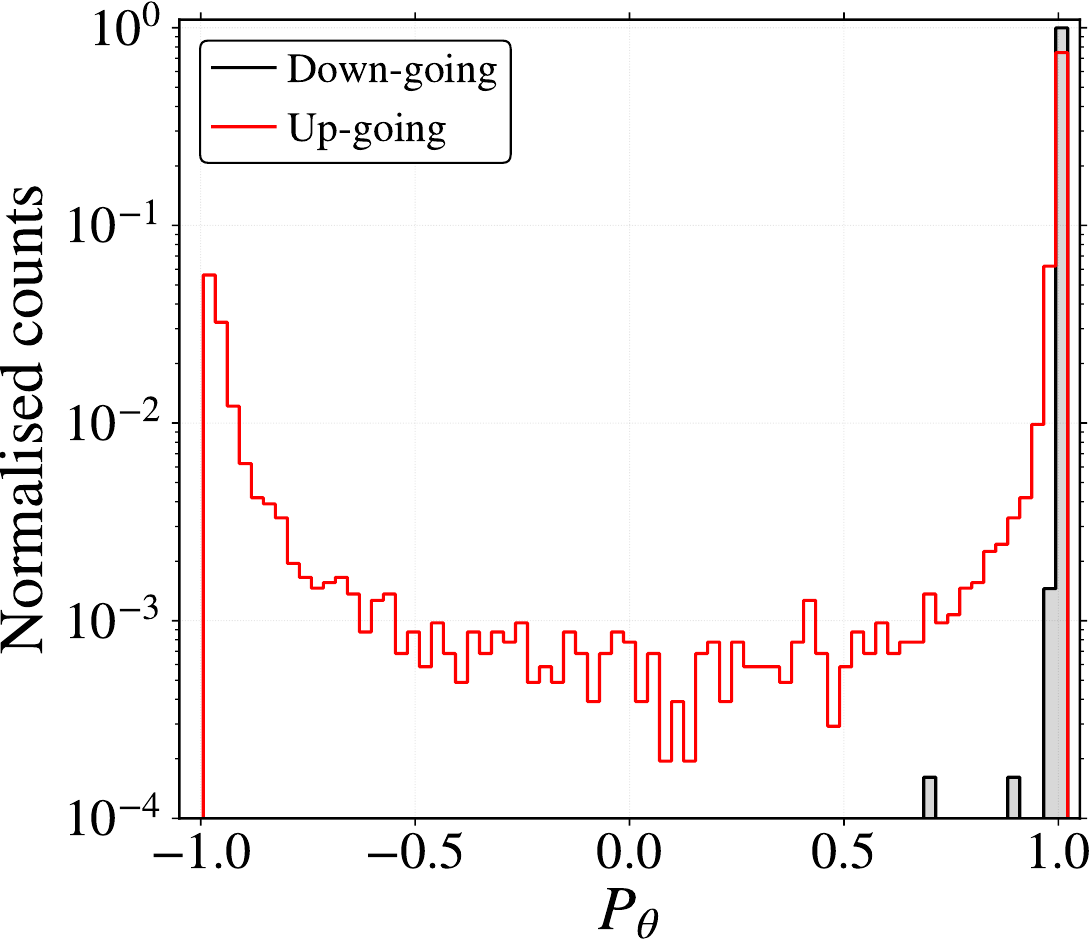}
\caption{Response of the Convolutional Neural Network (CNN) to upward-going and downward-going particles after the application of the cuts presented in Fig.~\ref{fig:scheme_model}. The up-going events correspond to electron- and muon-neutrinos while for the background there are single particles with different geometries ($1\,$GeV electrons, $2\,$GeV muons, $10\,$GeV protons and shower events induced by protons with energies $\in [100, 500]\,$GeV and $\theta \in [5^\circ,15^\circ]$.}
\label{fig:Ptheta} 
\end{figure}

The CNN was subject to comprehensive testing using independent datasets, each containing around $10^6$ events. These datasets included single particles, upward-going neutrino events and downward-going proton-induced showers, with energies from $100$ to $500\,$GeV. The results, as depicted in Fig.\,\ref{fig:Ptheta}, demonstrate that the network correctly identifies all background events, corresponding to $P_\theta > 0$, and all having $P_\theta > 0.5$ with a large peak at $P_\theta\simeq 1$. 
%

The small number of background events with $P_\theta < 0.8$ in Fig.\,\ref{fig:Ptheta} is primarily due to high-energy showers. The large particle multiplicity within the WCD makes it harder for the CNN network to confidently classify these showers as down-going events, because the CNN was not specifically trained for energetic shower events. However, it was confirmed in our simulations that for primary energies $\gtrsim 500\,$GeV, the footprint of proton-induced showers at the ground is $\sim 150\,$m wide. As such, the shower can trigger multiple WCD stations in the array allowing for effective vetoing.

\section{Neutrino Effective Mass}

To evaluate the proposed method for neutrino detection, we can estimate the effective detector mass available for upward-going neutrino identification using the array layout described in Section~\ref{sec:analysis}.

The effective mass depends on the direction $(\theta, \phi)$ and energy of incoming neutrinos ($E_{\nu}$), as well as on the position of their interaction point $(x,y,D)$, with $D$ defined as the depth (in ${\rm g\,cm^{-2}}$) measured from the WCD station at which neutrino interactions can be identified as up-going events. This quantity can be calculated, assuming a uniform distribution in $\phi$, using equations\,(\ref{eq:dEff_mass}) and (\ref{eq:Eff_mass_PointLike}). Equation\,(\ref{eq:dEff_mass}) allows to estimate the number of detected neutrinos under the assumption of an isotropic flux, while Eq.\,(\ref{eq:Eff_mass_PointLike}) assumes that the neutrinos have their origin in a point-like source.

\begin{widetext}
\begin{equation} \label{eq:dEff_mass}
    M_{\text{eff}} \ (E_{\nu}) = \int \dfrac{dM_{\text{eff}}}{d\theta} (\theta,E_{\nu}) \ {\rm d} \theta = 2\pi \ N_{\rm stations} \int \sin{\theta} \ \varepsilon(x,y,D,\theta,E_{\nu})\ {\rm d}x \ {\rm d}y \ {\rm d} D \ {\rm d}\theta \ \ {\rm [g]}
\end{equation}
\end{widetext}

\begin{equation}
M_{\text{eff}} \ (E_{\nu}, \theta) =  N_{\rm stations} \int \varepsilon(x,y,D,\theta,E_{\nu})\ {\rm d}x \ {\rm d}y \ {\rm d}D \ \ {\rm [g]}
\label{eq:Eff_mass_PointLike}
\end{equation}

It is important to note that in this approach, where only individual station units are used for neutrino detection, the effective mass is directly proportional to the number of WCD stations employed $N_{\rm stations}$.
The main element to be determined in Eqs.~(\ref{eq:dEff_mass}) and~(\ref{eq:Eff_mass_PointLike}) is the neutrino identification efficiency, $\varepsilon(x,y,D,\theta,\phi,E_{\nu}) \equiv {n_{\rm acc}}/{n_{\rm sim}} \in\left[0,1\right]$, where $n_{\rm acc}$ is the number of events passing the cuts described in Section~\ref{sec:analysis}, and $n_{\rm sim}$ the total number of simulated events. This number is obtained using simulated muon and electron neutrino-induced showers beneath the WCD with energies $E_0 = \left\{ 10\,{\rm GeV}, 1\,{\rm TeV}, 100\,{\rm TeV} \right\}$, and zenith angles $\theta = \left\{140^{\circ},150^{\circ},165^{\circ},180^{\circ}\right\}$.
The neutrino interaction point is sampled uniformly in an area bigger than the WCD projected area, and at different slant depths $D$ from the WCD station until the efficiency, $\varepsilon$, becomes negligible. About a hundred events were simulated for each neutrino interaction point. The effective mass is obtained by a numerical integration of Eqs.\,(\ref{eq:dEff_mass}) and (\ref{eq:Eff_mass_PointLike}), interpolating the discrete values of $\varepsilon(E_0,\theta,D)$~\cite{spline}. 

\begin{figure*}[htb]
 \centering
\includegraphics[width=0.75\linewidth]{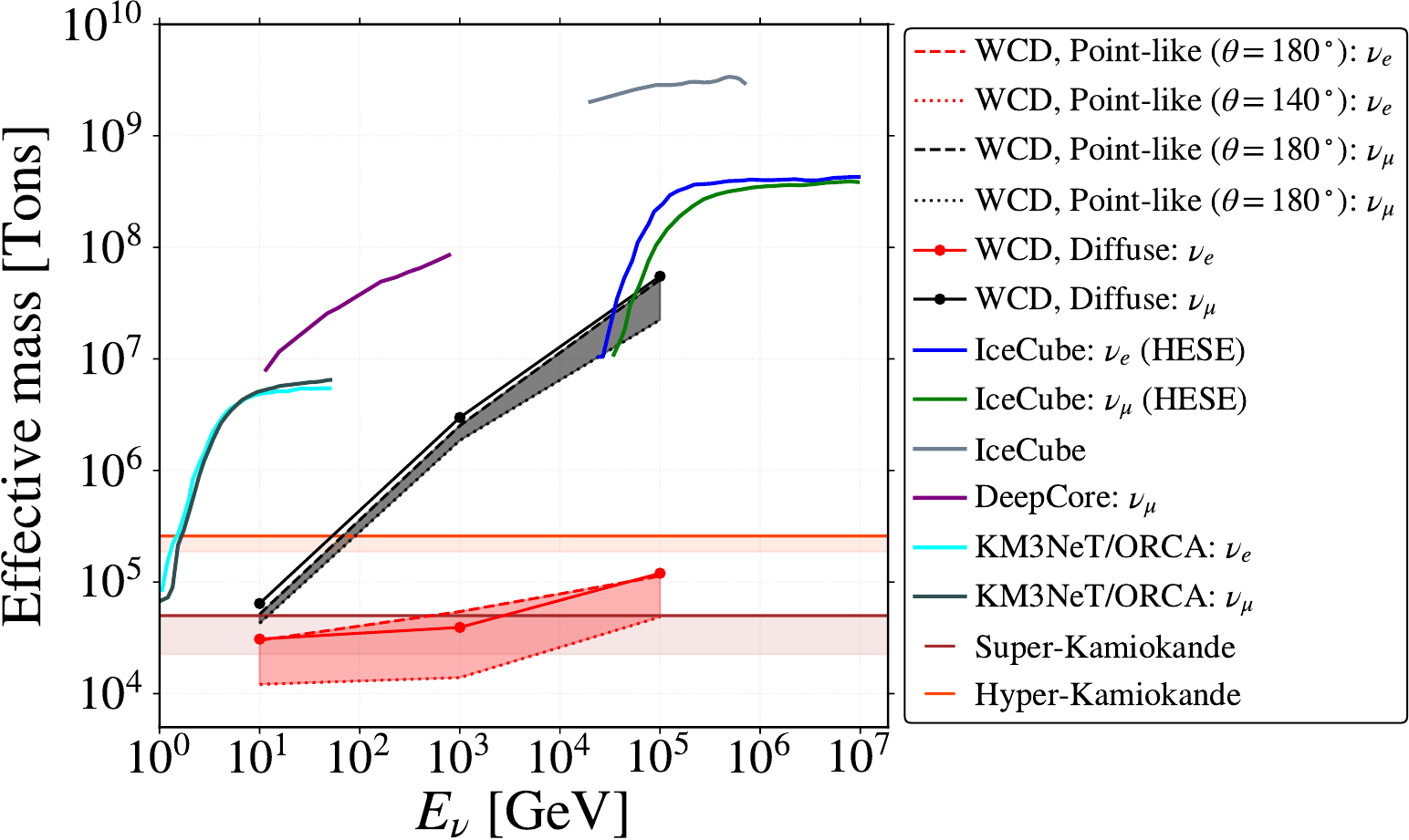}
\caption{Effective mass for a diffuse flux of neutrinos (points with lines) of the gamma-ray detector as described in Section\,\ref{sec:analysis} using its full array configuration. The effective masses of the dense and sparse arrays can be obtained scaling the effective mass of the total array by the factors F=$0.61$ and F=$0.39$ respectively. The shaded gray and red bands bracket the effective mass for a flux of neutrinos corresponding to a
point-like source (full gamma-ray detector array), with the upper edge of the bands (dashed lines) corresponding to vertical up-going events with $\theta = 180^\circ$, and the lower edge (dotted lines) to up-going events with $\theta = 140^\circ$. The results are compared with the effective masses of the following experiments: Super-Kamiokande \cite{SuperKamiokande}, Hyper-Kamiokande \cite{HyperKamiokande}, DeepCore online filter for $\nu_\mu$ (taken from Fig.\,2 of reference \cite{DeepCore2016}), KM3NeT/ORCA (taken from Fig.\,1 of reference \cite{KM3NeT2022EffMass}), IceCube (derived from Figures 2 and 5 of reference \cite{icecube2023observation}), and IceCube $\nu_e$ and $\nu_\mu$ High-Energy Starting Events (HESE) (taken from Fig.\,7B of reference \cite{icecube2013EffMass}). A shaded area between the fiducial and total water volume of the Kamiokande detectors was added. 
}
\label{fig:EffMass} 
\end{figure*}

The effective mass as a function of neutrino energy for the gamma-ray observatory described in Section \ref{sec:analysis}, and for the case of an isotropic neutrino flux, is shown in Fig.~\ref{fig:EffMass} (solid lines), together with the effective masses of other present and future neutrino observatories. A good sensitivity to atmospheric and astrophysical neutrinos can be achieved with this approach, complementing observations made by existing neutrino observatories. 
The effective mass for the case of a neutrino flux from a point-like source, is also shown in Fig.~\ref{fig:EffMass} (shaded bands). As expected, the effective mass is larger for vertically ascending trajectories (upper edge of the band corresponding to $\theta=180^\circ$) compared to more inclined ones (lower edge of band corresponding to $\theta=140^\circ$). The significantly larger effective mass of $\nu_\mu$ compared to $\nu_e$ in Fig.\,\ref{fig:EffMass}, both for the diffuse and point-like flux, arises from the fact that the interaction of $\nu_\mu$ with the Earth produces a high-energy muon capable of traversing large distances depending on its energy. 

It should be noted that, while some of the presented curves from other experiments include quality selection cuts, in this work we have not applied any extra quality cuts to enhance neutrino energy or direction reconstruction for obtaining the effective masses.

Finally, the approach explored in this work extends and complements the sky coverage of other experiments, enhancing the capabilities of a gamma-ray ground-based array of WCD at essentially no cost, and hence contributing to the multi-messenger network of detectors of cosmic particles.

\section{Discussion}

Our study has demonstrated that, with specific PMT configurations within a WCD station, it becomes feasible to efficiently discriminate particles moving in an upward-going trajectory, such as those in upgoing neutrino-induced showers interacting beneath the WCD, from those moving downward such those produced in conventional cosmic-ray atmospheric showers.

We tested the validity of our results by considering the electronic response characteristics of the PMTs~\cite{SWGO_detector} varying their sampling frequency from $1$ to $5\,$ns in the simulations. The response of the photosensors has been emulated using the tables provided in reference~\cite{Electronics}. These include the characterization of various aspects of the PMT and electronics response such as the single photoelectron (p.e.) amplitude distribution, pulse time distribution, and the electronics pulse shape.
Remarkably, in all the tests conducted, the distinguishing features in the PMT signal time trace that enabled the identification of up-going neutrinos remained visible. This observation holds true as long as the time sampling is kept below $5\,$ns.

\begin{figure}[htb]
 \centering
\includegraphics[width=0.35\textwidth]{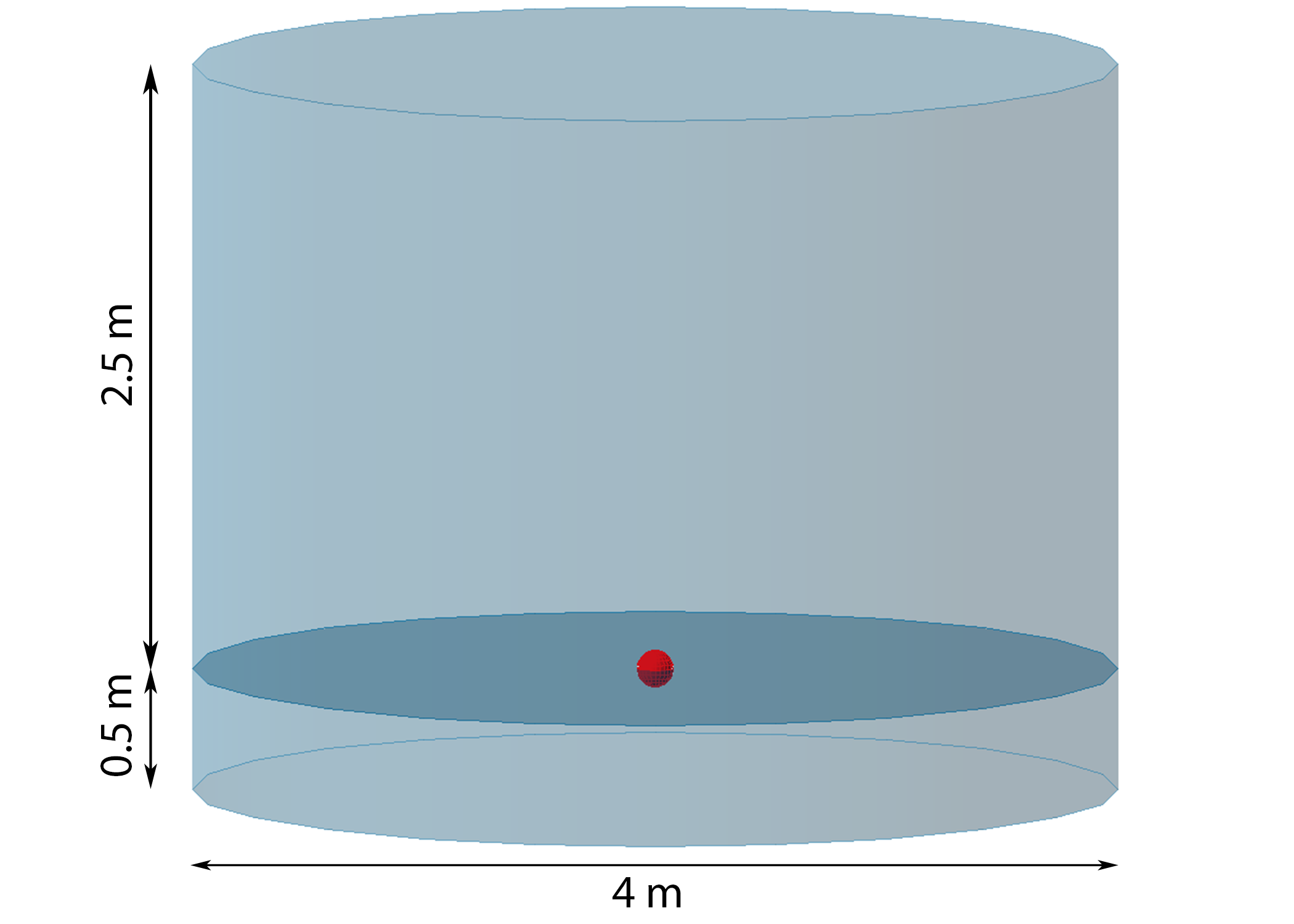}
 \caption{Illustration of a double-layered water-Cherenkov detector with two photosensors in each chamber. The two PMTs, one facing upwards and the other downwards, are represented by the red domes.}
 \label{fig:WCDs:DLWCD}
\end{figure}

The current study was carried out for a specific detector concept and array layout configuration. Nevertheless, our findings can be extrapolated to other experimental layouts as long as they also feature two photosensors separated by some vertical distance. For instance, we have checked that the methodology of applying simple cuts followed by a CNN, would also be very effective, after properly adjusting the cuts, in other setups, such as a double-layered WCD, as proposed in~\cite{DLWCD} and illustrated in Fig.~\ref{fig:WCDs:DLWCD}.

\begin{figure}[htb]
 \centering
\includegraphics[width=0.3\textwidth]{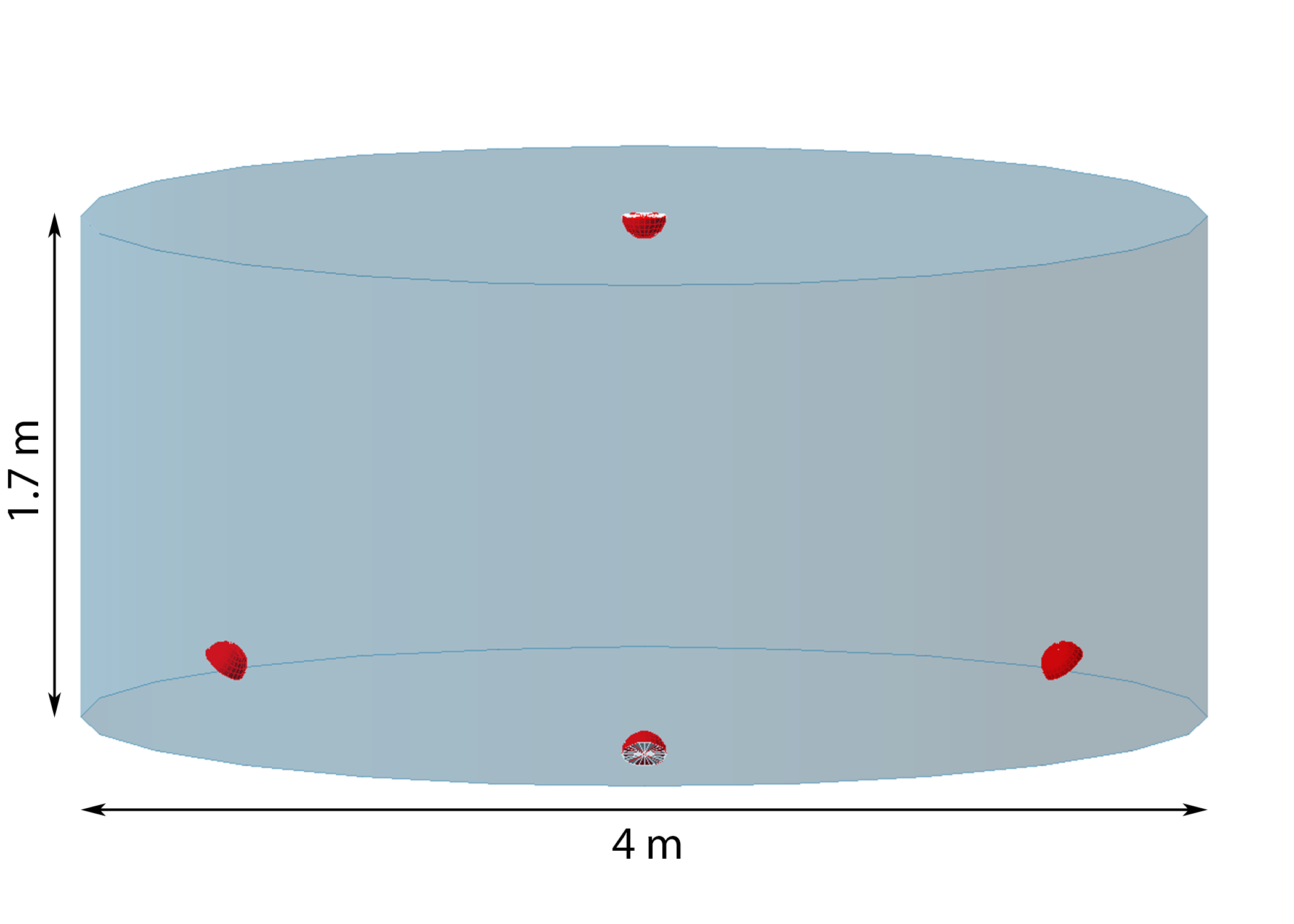}%
 \caption{Illustration of a water-Cherenkov detector with three photosensors at the bottom (\textit{Mercedes} configuration) and one at the top. The PMTs are represented by the red domes.}
 \label{fig:WCDs:M3T1}
\end{figure}

On the other hand, using only two PMTs, limits the accuracy in the reconstruction of the arrival direction of up-going neutrinos. Nevertheless, the use of multiple PMTs within a single station, as suggested in~\cite{wcd2022mercedes, WCD4PMTs}, can be explored to improve the accuracy of the angular reconstruction of upward-going neutrinos. WCDs with multiple PMTs have been proposed as an alternative for the SWGO compact array~\cite{WCD4PMTsICRC}.
For instance, using a WCD with 3 PMTs at the bottom (\textit{Mercedes} station) and one at the top, as sketched in Fig.\,\ref{fig:WCDs:M3T1}, a significant improvement in angular reconstruction can be achieved for up-going neutrino-induced showers of $1\,$TeV. For this particular setup, we carried out a dedicated study in which, the arrival direction of the incoming neutrino is reconstructed using two CNNs: one for determining its zenith angle, $\theta$, and another for the azimuthal angle, $\phi$. These two CNNs analyze the signal time traces from each PMT of events which have been tagged as up-going events, following the methodology illustrated in Fig.~\ref{fig:scheme_model}.
To enhance the angular reconstruction, the loss function of the neural networks was adapted to incorporate angles as observables. Specifically, a circular regression loss function of the form ${\rm Loss} = \frac{1}{2N} \sum^N_{i=1} [1 - \cos(y_i - \hat{y}_i)] \in [0,1]$ was used to ensure precise reconstruction within the range of 0 to 360 degrees, taking into account the periodicity of angles. Here, $N$ denotes the number of samples, $y_i$ represents the true angle for the $i$-th sample, and $\hat{y}_i$ denotes the predicted angle. The overall loss is computed as a sum across all samples, normalized by a factor of $1/2N$ to scale it within the range $[0,1]$, where zero indicates a perfect angular reconstruction. 

\begin{figure}[htb]
 \centering
\includegraphics[width=0.45\textwidth]{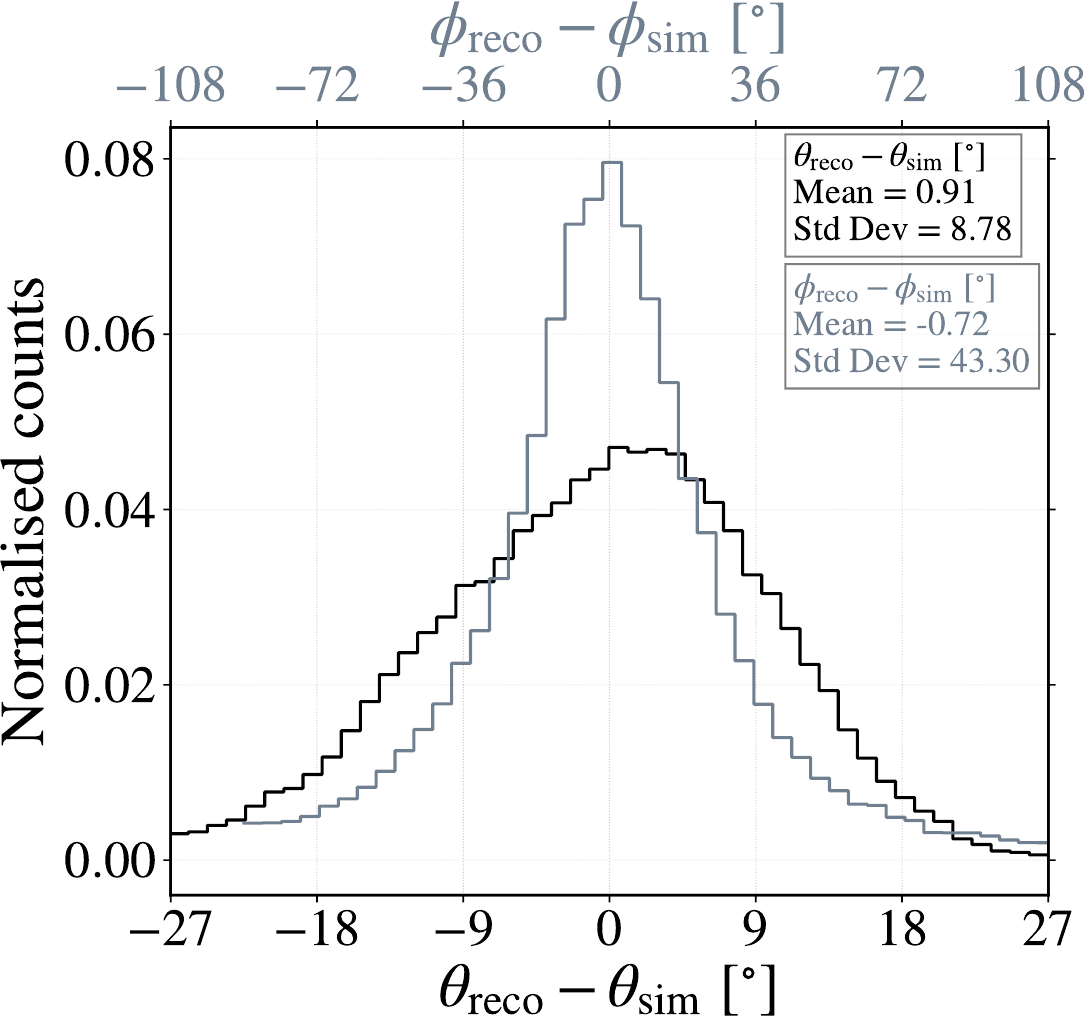}%
 \caption{Distributions of the difference between the reconstructed and simulated neutrino-induced shower azimuthal angle $\phi$ (gray and top axis) and zenith angle $\theta$ (black and bottom axis) using the M3T1 WCD \textit{Mercedes} design depicted in Fig.\,\ref{fig:WCDs:M3T1}. The width of the bins in the x-axis is 10$\%$ of the total range for the given angle $\theta$ or $\phi$.}
 \label{fig:geom_reco}
\end{figure}

The resolution in angular reconstruction is obtained as the standard deviation of the distribution of the difference between the reconstructed and simulated neutrino angles, as illustrated in Fig.~\ref{fig:geom_reco}.
A modest reconstruction resolution  of approximately $9^\circ$ in $\theta$ and approximately $43^\circ$ in $\phi$ was achieved, that would allow the determination of a relatively small solid angle in the sky, helping in the identification of potential transient neutrino sources.

Further improvement could be achieved in the future employing multiple smaller PMTs with equivalent photocathode-sensitive areas but oriented in various directions, as proposed in~\cite{hyperk,km3net,SWGO_detector}.

\begin{figure}[htb]
 \centering
  \includegraphics[width=0.45\textwidth]{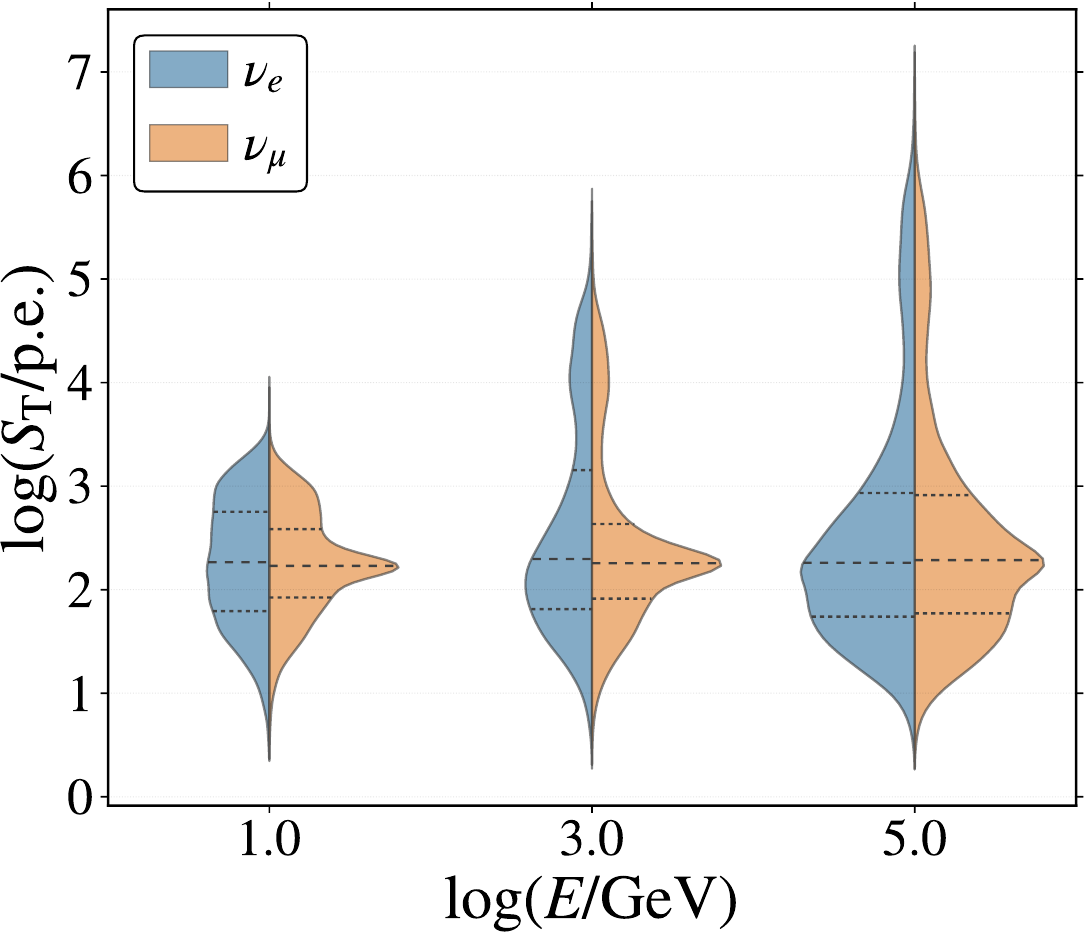}
 \caption{Distribution of the signal for electron and muon neutrino-induced showers. The shaded areas of the violin plot represent the normalised distribution of the total signal at the WCD for a given neutrino energy. The horizontal lines show the quartiles of the distributions. 
 }
 \label{fig:energy_reco}
\end{figure}

The resolution in the reconstruction of the neutrino energy is anticipated to be rather low, except in rare cases where high-energy neutrinos interact close to the detector, increasing the total recorded signal at the station (see Fig.~\ref{fig:energy_reco}). In essence, the WCD detector array can be regarded as a counter of neutrinos above a certain energy threshold.

Finally, we do not expect the site location to play an important role in the discrimination of upward from downward-going events. The key parameter is site altitude. As altitude increases, background from atmospheric muons and low-energy showers are expected to rise, though their energy spectrum changes very little. The methodology in this paper successfully handled atmospheric muons and vertical showers up to $1\,$TeV at $5200\,$m altitude, making the analysis strategy applicable to different altitudes.

\section{Conclusion}

The study presented in this work illustrates the potential of a single water-Cherenkov detector unit in identifying up-going neutrino-induced showers. 
To achieve this, the WCD must be equipped with at least two photosensors positioned at the bottom and the top of the station. A set of cuts on observables related to the signal and timing characteristics of the PMT signal traces, effectively removes the majority of background events when compared to upward-going events. However, due to some peculiar geometries of the shower particles traversing the WCD, a small fraction of $\lesssim 1\%$ of background events can only be rejected using a machine-learning algorithm trained on the features of the PMT signal time trace.

It has also been demonstrated that the arrival direction of upward-going neutrinos can be reconstructed with a resolution of approximately $9^\circ$ in $\theta$ and $43^\circ$ in $\phi$ for WCD stations equipped with 3 PMTs at the bottom and one at the top.
Although the resolution in energy reconstruction remains low even in this case, applying our methodology could turn an array of WCD stations devoted to gamma-ray detection into a viable counter of neutrinos above a certain energy threshold.

The findings presented in this paper can enhance the synergy between gamma/cosmic-ray and neutrino observatories in the search for transient sources of energetic cosmic particles in the universe.

\begin{acknowledgments}
The authors wish to express their appreciation for the financial support provided for this work by FCT - Fundação para a Ciência e a Tecnologia, I.P., through project PTDC/FIS-PAR/4300/2020 
and from Spain – Ministerio de Ciencia e Innovaci\'on/Agencia Estatal de Investigaci\'on
(PID2019-105544GB-I00, PID2022-140510NB-I00),
Xunta de Galicia (CIGUS Network of Research Centers,
Consolidaci\'on 2021 GRC GI-2033, ED431C-2021/22 and 2022 ED431F-2022/15),
and the European Union (ERDF).
B.S.G. (LIP/IST) is grateful for the financial support by the FCT PhD grant PRT/BD/151553/2021 under the IDPASC program. 
\end{acknowledgments}

\bibliography{references}



\end{document}